\documentstyle[aps]{revtex}

\begin{document}
\draft
\title{Single and vertically coupled type II quantum dots in a perpendicular
magnetic field: exciton groundstate properties}
\author{K. L. Janssens\cite{karenmail}, B. Partoens\cite{bartmail} and F. M. Peeters
\cite{peetersmail}}
\address{Departement Natuurkunde, Universiteit Antwerpen (UIA), Universiteitsplein 1,%
\\
B-2610 Antwerpen, Belgium}
\date{\today}
\maketitle

\begin{abstract}
The properties of an exciton in a type II\ quantum dot are studied
under the influence of a perpendicular applied magnetic field. The
dot is modelled by a quantum disk with radius $R$, thickness $d$
and the electron is confined in the disk, whereas the hole is
located in the barrier. The exciton energy and wavefunctions are
calculated using a Hartree-Fock mesh method. We distinguish two
different regimes, namely $d<<2R$ (the hole is located at the
radial boundary of the disk) and $d>>2R$ (the hole is located
above and below the disk), for which angular momentum $(l)$
transitions are predicted with increasing magnetic field. We also
considered a system of two vertically coupled dots where now an
extra parameter is introduced, namely the interdot distance
$d_{z}$. For each $l_{h}$ and for a sufficient large magnetic
field, the ground state becomes spontaneous symmetry broken in
which the electron and the hole move towards one of the dots. This
transition is induced by the Coulomb interaction and leads to a
magnetic field induced dipole moment. No such symmetry broken
ground states are found for a single dot (and for three vertically
coupled symmetric quantum disks). For a system of two vertically
coupled truncated cones, which is asymmetric from the start, we
still find angular momentum transitions. For a symmetric system of
three vertically coupled quantum disks, the system resembles for
small $d_{z}$ the pillar-like regime of a single dot, where the
hole tends to stay at the radial boundary, which induces angular
momentum transitions with increasing magnetic field. For larger
$d_{z}$ the hole can sit between the disks and the $l_{h}=0$ state
remains the groundstate for the whole $B$-region.
\end{abstract}


\section{Introduction}

The study of self-assembled quantum dots, as realized by the
Stranski-Krastanow growth mode, has been a fascinating research
area during the last decade. The possibility to discover new
physics in these zero-dimensional (0D) structures together with
possible applications in opto-electronics has led to many
experimental and theoretical results. Most studies were devoted to
type I quantum dots\cite
{polimeni,wilson,stier,brasken,bayer,ulloa,karen,korkusinski},
where both electron and hole are spatially located inside the
quantum dots. Relatively few works however were done for type II
quantum dots, where electrons and holes are spatially separated.
One expects from this type of dots interesting properties, such as
large tunability of the emission energy and radiative lifetimes
which are considerably longer than for their type I
counterparts\cite{sun}.

Within the group of type II quantum dots, one can still
distinguish between two systems. Namely, the system which confines
the electrons inside the quantum dot, and forms an antidot for the
hole, like e.g. the $InP/GaInP$ dots (if strain is neglected).
Besides this, there are the systems, such as $GaSb/GaAs$ \cite
{muller-kirsch} and $InAs/${\it Si} \cite{heitz}, where the holes
are confined within the dot, but where the electrons are located
outside. Experimental studies on this type of quantum dots have
been performed in Refs. \cite{nomura,hayne,sugisaki} for the
$InP/GaInP$ system and in Refs. \cite{muller,hatami} for the
$GaSb/GaAs$ system. Theoretical investigations on $InP/GaInP$ type
II dots were performed by Pryor {\it et al}. \cite{pryor} and
Tadic {\it et al.} \cite{milan}, who\ used a strain-dependent
${\bf k}\cdot {\bf p}$ Hamiltonian to calculate the electronic
structure. Nomura {\it et al}. \cite{nomura2} calculated the
Landau levels in a high magnetic field, solving the Hartree
equations self-consistently, while the present authors studied an
exciton in a planar quantum dot using a Hartree-Fock approximation
\cite{karen2}. Lelong {\it et al}. \cite{lelong} calculated the
binding energy of excitons, charged excitons and biexcitons in
$GaSb/GaAs$
dots using the Hartree-Fock approximation. The magneto-exciton in a $%
GaSb/GaAs$ dot was investigated by Kalameitsev {\it et al}.\ \cite
{kalameitsev}.

In the present paper we study a single electron-hole pair bound by
the Coulomb interaction, i.e. an exciton, in a {\it model} type II
quantum dot. The quantum dot is modelled by a quantum disk of
finite height, which is an extension of our previous work
\cite{karen2} where we did not take into account the $z$
-direction. As an example, we consider the electron in the dot and
the hole in the barrier material. The conclusions of our paper are
also valid for the reverse situation where the hole is inside the
dot and the electron is outside the dot. In the present work, we
neglect strain effects and for the confinement potential we use a
hard wall of finite height, confining the electron inside the dot
and repelling the hole to the barrier region. Furthermore, a
magnetic field was applied in the growth direction, i.e. ${\bf
B}=B{\bf e}_{z}$.

The first part of this paper deals with the study of the exciton
in a single dot. Investigation of the influence of the disk
parameters, namely the radius $R$ and thickness $d,$ showed that
one can distinguish between two regimes: the disk-like regime and
the pillar-like regime. For the first one, the disk thickness $d$
is much smaller than its diameter, i.e. $d<<2R,$ whereas the
latter one describes the system with $d>>2R$. Although the
magnetic field couples only with the in-plane electron and hole
coordinates, we will show that these systems, with different
$z$-extension, behave very differently under an applied magnetic
field.

In the second part of the paper, we study the properties of
vertically coupled quantum dots. The interest of this study lies
in the following. It is generally known that in reality
self-assembled quantum dots resemble more our disk-like system
than our pillar-like system. However, it is also known that it is
not too difficult to form vertical stacks of these disk-like dots
and thereby create a system which could behave like a pillar-like
system.

The paper is organized as follows. In Sec. II, we give a brief
discussion of our theoretical model. The numerical results in the
absence of a magnetic field are presented in Sec. III. Section IV
deals with the result for a perpendicular applied magnetic field.
Part A of this Section discusses the results for the single dot
disk-like system, whereas Part B is dedicated to the pillar-like
system. Parts C and D deal with two vertically coupled dots,
respectively disks and truncated cones. Parts E and F deal with
three vertically coupled dots, respectively for small and large
interdot distances. Finally, we make a small revision of the
results for the single disk in Part G. Our results are summarized
in Sec. V. In the Appendix, we discuss in more detail the
numerical approach we used for the calculation of the Hartree
integral.

\section{Theoretical model}

We extended our previous approach \cite{karen2}, which was valid
for planar dots, and include the $z$-direction. The energies and
wavefunctions are solved within the effective mass approximation
(with $m_{e}$ and $m_{h}$ the effective electron and hole masses,
respectively, $r_{e,h}=\sqrt{x_{e,h}^{2}+y_{e,h}^{2}},$ $\omega
_{c,e}=eB/m_{e}$ and $\omega _{c,h}=eB/m_{h}$) and the
Hartree-Fock (HF) single particle equations can be written as
\begin{eqnarray}
\left[ -\frac{\hbar ^{2}}{2m_{e}}\frac{1}{r_{e}}\frac{\partial }{\partial
r_{e}}\left( r_{e}\frac{\partial }{\partial r_{e}}\right) -\frac{\hbar ^{2}}{%
2m_{e}}\frac{\partial ^{2}}{\partial ^{2}z_{e}}+\frac{\hbar ^{2}}{2m_{e}}%
\frac{l_{e}^{2}}{r_{e}^{2}}+\frac{l_{e}}{2}\hbar \omega _{c,e}+\frac{1}{8}%
m_{e}\omega _{c,e}^{2}r_{e}^{2}\right. &&  \nonumber \\
\left. +V_{e}(r_{e},z_{e})-\frac{e^{2}}{4\pi \epsilon }\int \frac{\rho
_{h}(r^{\prime },z^{\prime })}{|{\bf r}-{\bf r}^{\prime }|}d{\bf r}^{\prime }%
\right] \psi _{e}(r_{e},z_{e})=\epsilon _{e}\psi _{e}(r_{e},z_{e}), &&
\eqnum{1a} \\
\left[ -\frac{\hbar ^{2}}{2m_{h}}\frac{1}{r_{h}}\frac{\partial }{\partial
r_{h}}\left( r_{h}\frac{\partial }{\partial r_{h}}\right) -\frac{\hbar ^{2}}{%
2m_{h}}\frac{\partial ^{2}}{\partial ^{2}z_{h}}+\frac{\hbar ^{2}}{2m_{h}}%
\frac{l_{h}^{2}}{r_{h}^{2}}-\frac{l_{h}}{2}\hbar \omega _{c,h}+\frac{1}{8}%
m_{h}\omega _{c,h}^{2}r_{h}^{2}\right. &&  \nonumber \\
\left. +V_{h}(r_{h},z_{h})-\frac{e^{2}}{4\pi \epsilon }\int \frac{\rho
_{e}(r^{\prime },z^{\prime })}{|{\bf r}-{\bf r}^{\prime }|}d{\bf r}^{\prime }%
\right] \psi _{h}(r_{h},z_{h})=\epsilon _{h}\psi _{h}(r_{h},z_{h}), &&
\eqnum{1b}
\end{eqnarray}
\noindent where we made use of the axial symmetry by taking $\Psi
_{e}(r_{e},\varphi _{e},z_{e})=e^{il_{e}\varphi _{e}}\psi
_{e}(r_{e},z_{e})$ and $\Psi _{h}(r_{h},\varphi
_{h},z_{h})=e^{il_{h}\varphi _{h}}\psi _{h}(r_{h},z_{h}),$ and
where the densities $\rho _{e}(r^{\prime },z^{\prime })$ and $\rho
_{h}(r^{\prime },z^{\prime })$ are given by respectively $\left|
\Psi _{e}(r_{e},\varphi _{e},z_{e})\right| ^{2}$ and $\left| \Psi
_{h}(r_{h},\varphi _{h},z_{h})\right| ^{2}$. We made use of a
finite difference scheme to solve the Hartree-Fock equations. More
details about the implementation of this finite difference scheme
can be found in Refs.~ \cite{peeters,karen2}. Note that we
consider only a single electron and single hole and therefore
there are no exchange terms. The reason that these equations can
still be called HF lies in the fact that the self-interaction is
excluded. As confinement potentials we take hard walls of finite
height:
\begin{equation}
V_{e,h}(r_{e},z_{e},r_{h},z_{h})=\left\{
\begin{array}{ll}
V_{e,h}, & r_{e,h}>R\text{ and\ }\left| z_{e,h}\right| >d/2, \\
0,&\mbox{otherwise},
\end{array}
\right.  \eqnum{2}
\end{equation}
with $R$ the radius of the disk, and where we took $V_{e}$
positive and $V_{h}$ negative. Note that the only good quantum
number is the total angular momentum in the $z$-direction, defined
by $L=l_{e}+l_{h}.$

We solved the equations self-consistently using an iterative procedure.
Since only the electron is confined in the absence of any Coulomb
interaction, we start with the free electron solution. The Hartree integrals
are then integrated numerically as
\begin{equation}
\int \frac{\rho (r^{\prime },z^{\prime })}{|{\bf r}-{\bf r}^{\prime }|}d{\bf %
r}^{\prime }=4\int dz^{\prime }\int dr^{\prime }\frac{\rho (r^{\prime
},z^{\prime })r^{\prime }}{\sqrt{\left( r+r^{\prime }\right) ^{2}+\left(
z-z^{\prime }\right) ^{2}}}{\cal K}\left( \frac{4rr^{\prime }}{(r+r^{\prime
})^{2}+\left( z-z^{\prime }\right) ^{2}}\right) ,  \eqnum{3}
\end{equation}
where ${\cal K}(x)$ is the complete elliptic integral of the first kind.
More details about the calculation and numerical implementation of this
integral is given in the Appendix.

After convergence of the iteration procedure, the total energy is given by
\begin{equation}
E_{\mbox{exciton}}=\epsilon _{e}+\epsilon _{h}+\frac{e^{2}}{4\pi \epsilon }%
\int \int \frac{\rho _{e}(r,z)\rho _{h}(r^{\prime },z^{\prime })}{|{\bf r}-%
{\bf r}^{\prime }|}d{\bf r}d{\bf r}^{\prime }.  \eqnum{4}
\end{equation}
The contribution of the correlation to the total energy is neglected in HF,
but for the self-assembled quantum dots, it is expected to be less than 2\%
\cite{brasken} and for type II dots this will be even less.

\section{Numerical results in the absence of an applied magnetic field}

In our model system, there are two main parameters which can be varied: the
radius of the disk $R,$ and the thickness of the disk $d.$ In the first part
of our numerical study we will investigate the influence of these parameters
on the exciton energy and wavefunction in the absence of a magnetic field.
Following material parameters, typical for the $InP/InGaP$ dot system, were taken: $m_{e}=0.077m_{0},$ $%
m_{h}=0.60m_{0},$ $\epsilon =12.61,$ $V_{e}=250meV$ and $V_{h}=-50meV.$

The variation of the disk radius $R$ and thickness $d$ has large
consequences for the hole wavefunction. Namely, the hole always
tends to sit as closely as possible to the electron. In this way,
when the disk thickness is larger than its diameter $\left(
d>>2R\right) $, the hole will prefer to sit at the radial boundary
of the disk. When we consider very thin disks however $\left(
d<<2R\right) $, it will be much more favourable for the hole to be
located above and below the disk. This state was not possible in
our previous study \cite{karen2} of the two-dimensional flat
quantum dot, where the hole was forced to sit at the radial
boundary of the dot.\ To illustrate this, we calculated the
probability to find the hole at the radial boundary of the disk as
\begin{equation}
P_{side}=2\pi \int_{-\infty }^{\infty }dz_{h}\int_{R}^{\infty
}dr_{h}~r_{h}\left| \Psi _{h}(r_{h},z_{h})\right| ^{2}.  \eqnum{5}
\end{equation}
When calculating this value for varying $R$ and $d,$ we obtained a
phase diagram for the position of the hole wavefunction, which is
shown in Fig.~1(a). The hole confinement potential in all these
calculations was fixed to $V_{h}=-50meV.$ The solid curve
indicates where 50\% of the hole is located at the radial
boundary, for the dashed curves it is respectively 25\% and 75\%.
In order to have a more visual picture of the hole state, we
present contourplots of the hole wavefunction $\left| \Psi
_{h}(r_{h},z_{h})\right| ^{2}$ for three characteristic situations
in the phase diagram, as indicated by the numbered stars.
Fig.~1(b) shows the hole density for $R=4nm,$ $d=12nm,$ where the
dotted lines indicate the boundary of the quantum disk. We clearly
see that the hole is mainly situated at the radial side of the
disk. Fig.~1(c) was made for $R=8nm,$ $d=8nm$ and at this
position, exactly 50\% of the wavefunction is located at
$r_{h}>R$. The third plot, Fig.~1(d), depicts the result for a
thin disk, with $R=12nm$ and $d=4nm,$ where the hole wavefunction
is mostly located above and below the disk, where it is nearer to
the electron.

Thus we can distinguish between two main systems: a) a disk-like
system with $d<<2R,$ where the hole will be located above or below
the quantum disk, and b) a pillar-like system with $d>>2R,$ where
the hole sits mainly at the radial boundary of the disk. The solid
line in Fig.~1(a) can be approximated by the curve
$d=1.76R-5.41nm.$ We will see in the following section that a
magnetic field has a very different effect on these two classes of
systems.

\section{Results for a perpendicular applied magnetic field}

\subsection{Single dot: a disk-like system}

After having discussed the influence of the main parameters of our
system, we are now ready to consider an extra feature, i.e. a
magnetic field applied along the $z$-direction, parallel to the
growth direction. The magnetic field will squeeze the
wavefunctions in the radial direction. When $d<<2R,$ the hole is
above and below the disk, and will be pressed stronger to the
center of the disk. Fig.~2 depicts the exciton energy as a
function of the magnetic field for a disk with $R=10nm$ and
$d=2nm$ for two different values of the hole angular momentum
$l_{h}.$ Note that this exciton energy includes both the one
particle energies and the Coulomb energy, i.e. $%
E_{exciton}=E_{e}+E_{h}-E_{coul}.$ In the present case the single hole
energy $E_{h}$ will be 0, as the single hole is not confined. The single
electron energies at $B=0T$ and $B=50T$ are given by respectively $%
E_{e}=192.6meV$ and $E_{e}=212.3meV.$\ We find a strong
enhancement of the exciton energy of almost $20meV$ for
$B=0T\rightarrow B=50T.$ Due to the magnetic field, the electron
and hole are both pushed closer to the center, which enhances the
Coulomb interaction. However, also the one particle electron
energy increases drastically with increasing magnetic field, thus
cancelling the effect of the enhanced (negative) Coulomb energy.
The two insets show what happens with the electron (dashed curves)
and hole (full curves) wavefunctions when a magnetic field is
applied. For $B=0T$ (Fig.~2(a)) the electron is strongly confined
in the quantum disk, and the hole is localized above and below the
disk. From the figure, it is obvious that this position is
energetically most favourable for the hole, since the hole will be
very close to the electron. Because of the very small thickness,
there is some smearing of the electron wavefunction out of the
disk in the $z$-direction, which attracts the hole to this
position. The corresponding wavefunctions for $B=50T$ are plotted
in Fig.~2(b), where again the dashed curve indicates the electron
wavefunction, and the solid curve the hole. We see that both
particles are compressed in the radial direction. For the
electron, this leads to a further penetration of the wavefunction
into the barrier. Because of this, the hole will be even stronger
attracted to the top and bottom of the disk, which leads to a
larger penetration of the hole in the disk. As a result, the hole
wavefunction will also be slightly squeezed in the $z$-direction.

\subsection{Single dot: a pillar-like system}

The exciton energy as a function of the magnetic field is plotted in Fig.~3
for the case of $R=4nm,$ $d=12nm$ and $V_{h}=-50meV.$ An interesting feature
which appears is the occurrence of (hole) angular momentum transitions with
increasing magnetic field. The origin of the angular momentum transitions is
the following.\ As discussed in Sec. III, the preferred position of the hole
for a disk with a thickness larger than its diameter will be at the radial
boundary of the disk. When a magnetic field is applied along the $z$%
-direction, the hole will be pushed against the border of the disk. Since
the disk forms a barrier for the hole, it will be energetically more
favourable to jump to a higher angular momentum $l_{h}$ state. Notice that $%
l_{h}$ is an approximate quantum number. This feature was also
apparent in our previous study of planar type II\ dots
\cite{karen2}, where we forced the hole to sit at the radial
boundary. This is similar to an exciton in a quantum-ring
structure \cite {ulloa2,govorov}. In the present study however, we
allow for the hole to relax also in the $z$-direction. Therefore,
for a small thickness of the disk (disk-like system), it will be
much more favourable for the hole to move in the $z$-direction and
only for thick enough disks, such $l_{h}$ transitions occur.
Contourplots of the wavefunctions are shown, respectively for the
electron at $B=0T$ (inset of Fig.~3) and for the hole at $B=0T,$
$B=25T,$ $B=50T$ and $B=90T$ (Figs.~4(a)-(d)). We find that the
electron is located in the center of the disk, with almost no
penetration into the barrier. At $B=0T,$ the hole is mainly
sitting at the radial boundary of the disk, although there is some
extent of the wavefunction towards the top and bottom of the disk,
and there is also some penetration into the dot. When increasing
the magnetic field, the hole jumps to the $l_{h}=1$ state around
$B\simeq 15T.$ We plotted the hole wavefunction at $B=25T$
(Fig.~4(b)), and we see that the hole is pushed at the side, where
it has now more space to extend in the radial direction. Further
increasing the magnetic field, leads to a jump to $l_{h}=2$ at
$B\simeq 45T$ and to $l_{h}=3$ at $B\simeq 79T.$ The contourplots
were made for respectively $B=50T$ (Fig.~4(c)) and $B=90T$
(Fig.~4(d)), and we see again that the hole is sitting at the side
of the disk. Notice that, for all four plots, the probability to
find the hole at $r\simeq 12nm$ is almost the same, i.e. $\left|
\Psi _{h}(r_{h}=12nm,z_{h}=0)\right| ^{2}\simeq 0.025.$ In the
absence of those $l_{h}$ transitions, the magnetic field would
have compressed the hole much more strongly to the radial boundary
of the disk which leads to a larger energy as is shown clearly by
the solid curve in Fig.~3.

In order to investigate the influence of the hole confinement
potential $V_{h}$ on the angular momentum transitions, we made a
phase diagram of the angular momentum state of the hole as a
function of both the confinement potential and the magnetic field.
This is shown in Fig.~5 for a disk with radius $R=6nm$ and
thickness $d=14nm.$ The figure shows that up to $V_{h}=-13meV$ no
angular momentum transitions occur. In this region $V_{h}$ is too
small to form a barrier for the hole, and the hole jumps inside
the disk due to the Coulomb interaction, forming a type I system.
We also see that the transition between type I and type II is very
sharp. In our previous study of the planar dots \cite{karen2}, we
found from a similar figure a re-entrant behaviour to the
$l_{h}=0$ state for sufficient large magnetic fields. This is not
so evident in the present case, as we find that the line which
separates the type I from the type II behaviour is almost
vertical.

From an experimental point of view, it is interesting to look at
the probability for recombination of the exciton, which is
proportional to the square of the overlap integral
\begin{equation}
I=\int \Psi _{e}({\bf r})\Psi _{h}({\bf r})d{\bf r=}\int_{0}^{2\pi
}e^{i(l_{e}+l_{h})\varphi }d\varphi \int_{0}^{\infty }\psi
_{e}(r)\psi _{h}(r)rdr. \eqnum{6}
\end{equation}
The integral over the angle gives $2\pi \delta _{l_{e}+l_{h}}.$
This means that the probability for de-excitation is only non-zero
for $l_{e}+l_{h}=0.$ This implies that, after an angular momentum
transition, the probability for recombination decreases strongly,
leading to a vanishing of the photoluminescence (PL) spectrum
after a certain value of the magnetic field. Fig.~6 depicts the
overlap integral $I$ as a function of the confinement potential of
the hole, for both $B=0T$ (solid curve) and $B=25T$ (dashed
curve). For $B=0T$ the ground state will always be the $l_{h}=0$
state, and the overlap integral will always be non-zero. However,
as the system goes from type I to type II we see a strong decrease
of the overlap, because the exciton becomes spatially indirect.
For $B=25T$ we find a different picture: after the system has
become type II, the overlap integral becomes zero. This is due to
the fact that at $B=25T$ an angular momentum transition has
occurred and the condition $l_{e}+l_{h}=0$ is no longer fulfilled.

\subsection{Two vertically coupled dots}

The inclusion of the $z$-direction in our calculation enables us
to investigate systems of vertically coupled quantum dots. We have
now an extra parameter to vary, namely the interdot distance
$d_{z}$. In this first part, we focus on two stacked dots of the
same size, namely radius $R=6nm$ and thickness $d=6nm$, and with
$d_{z}=3.6nm$. The total stack height of $15.6nm$ is thus larger
than the diameter of the disks, being $12nm$. Therefore we expect
to find a magnetic field behaviour which resembles that of the
single disk pillar-like system. The result for the exciton energy
as function of the magnetic field is shown in Fig.~7. Indeed we
find angular momentum transitions with increasing magnetic field.
The inset, Fig.~7(a), shows a contourplot of the electron
wavefunction at $B=0T$, which is symmetrically distributed over
the two disks. The Coulomb interaction attracts the hole as close
as possible to the electron, in this case the radial boundary of
the disks. As one can see in Fig.~7(b), the hole tends also to sit
between the disks, to be even closer to the electron. A magnetic
field acts on the wavefunctions in the radial direction, and
thereby pushes the hole stronger between the disks. There is
however not enough space, and therefore it will be energetically
more favourable for the hole to jump to a higher $l_{h}$-state.

When giving a closer look at the behaviour of the different
$l_{h}$-states in a magnetic field (Fig.~7), a remarkable feature
appears: the $l_{h}=0$ curve exhibits a kind of kink around
$B=15T$, and we find a similar feature for $l_{h}=1$ around
$B=45T$. Investigation of the wavefunctions (Figs.~8(a-c)) learns
us that we are dealing with a \textit{spontaneous symmetry
breaking}\cite{uitleg}. The corresponding hole wavefunction at
$B=15T, 30T$ and $50T$, for $l_{h}=0$ is shown in Fig.~8(a,b,c)
respectively, where one can see clearly the increase of asymmetry
with magnetic field. But for such magnetic fields the $l_{h}=0$
state is not the ground state. This means that the jump to a
higher angular momentum state is still preferred above the
asymmetric state (see Fig.~8(d) for $l_{h}=3$ and $B=50T$). We
also found that with increasing $l_{h}$ the symmetry breaking will
occur at larger magnetic fields.

In order to investigate whether or not the ground state
configuration is asymmetric, we calculated
\begin{equation}
\Delta P_{asymm}=2\pi \int_{0}^{\infty }dr_{h}~r_{h} \left(
\int_{0}^{\infty }dz_{h}-\int_{-\infty }^{0}dz_{h}\right) \left|
\Psi _{h}(r_{h},z_{h})\right| ^{2}, \eqnum{7}
\end{equation}
which expresses the degree of asymmetry of the hole wavefunction
in the $z$-direction. This quantity is plotted in Fig.~9 as
function of the magnetic field for $l_{h}=0,1,2,3$. This plot
confirms that after $B \simeq 15T$ the $l_{h}=0$ state becomes
highly asymmetric, and for higher $B$ also $l_{h}=1$ becomes
asymmetric. The asymmetric states for the subsequent $l_{h}$
states occur for $B > 50T$. The contourplot of the hole
wavefunction at $B=50T$ is shown in Fig.~8(d) for the ground
state, which is $l_{h}=3$. The inset of Fig.~9 shows the dipole
moment for the different $l_{h}$ states with increasing magnetic
field. This is an interesting quantity, as it can be measured
experimentally. Note that for our symmetric system, we have the
unique feature that a magnetic field is able to induce a dipole
moment!

We can understand the spontaneous symmetry breaking as follows. In
the asymmetric case the electron wavefunction is of course
attracted to the hole (see Fig. 8(b), dashed lines) and thus
stronger confined, in this way increasing its single particle
energy. Therefore, the broken symmetry state can only occur if it
can overcome this increase in energy due to a larger Coulomb
attraction between electron and hole. We mentioned that an
increasing magnetic field favoured the broken symmetry state. The
reason is that with increasing magnetic field the electron becomes
stronger confined in the dot, and consequently expells more in the
$z$-direction. In the broken symmetry case it even expells more,
resulting in a larger overlap with the hole, which can eventually
make the asymmetric state lower in energy than the symmetric
state. If this is true, than the broken symmetry would occur even
sooner with increasing magnetic field for thinner disks, where the
electron is expelled more in the barrier material in the
$z$-direction. That this is indeed the case is shown in Fig.~10
where for two thin vertically coupled disks ($R=12$ nm, $d=3$ nm
and $d_z=$3nm, $l_h=0$) we find already a broken symmetry state
for zero magnetic field. In this case, this broken symmetry state
is also the ground state.

\subsection{Two vertically coupled truncated cones}

It is now interesting to check whether the above angular momentum
transitions survive in a system which is completely asymmetric
from the beginning, such as a system of two vertically coupled
truncated cones. The two cones are equal, and defined by a base
radius $R_{b}$ of $7nm$, a thickness $d=6nm$, and an angle of
$71^{\circ}$ (dashed lines in the insets of Fig.~11). The hole and
electron wavefunctions (Figs.~11 (a) and (b)) show indeed the
asymmetric behaviour, at $B=0T$ and for $l_{h}=0$. The result for
the exciton energy as a function of the magnetic field is depicted
in Fig.~11. The interesting feature is that, even for this system,
angular momentum transitions occur. Since there is still some part
of the hole wavefunction located at the radial border and also
some part between the disks, it appears to be still more
favourable for the hole to jump to the higher $l_{h}$-state than
to move more and more below the disk. Figs.~12(a-c) show
contourplots of the hole wavefunction in the ground state for
subsequent $l_{h}$-states, respectively $l_{h}=1$ at $B=20T$,
$l_{h}=2$ at $B=30T$, and $l_{h}=3$ at $B=50T$.

\subsection{Three vertically coupled dots:\ small interdot distance $d_{z}$}

For simplicity we consider only the case of three identical dots.
The first system under investigation contains three vertically
coupled quantum disks, each with disk radius $R=8nm,$ disk
thickness $d=3nm,$ and with interdot distance $d_{z}=3nm.$ In this
case the confinement of the hole $V_{h}$ was taken to be $-30meV.$
Fig.~13 shows the result for the exciton energy as a function of
the magnetic field. Similar to the case of the pillar-like single
dot, we find angular momentum transitions with increasing magnetic
field. The origin of these angular momentum transitions can be
understood by looking at the wavefunctions. The insets of Fig.~13
show the electron wavefunctions at (a) $B=0T$ and (b) $B=50T$, and
it appears that the main part is located in the middle dot,
whereas there is some small extent of the wavefunction into both
the upper and lower dots. At $B=50T,$ the wavefunction is more
squeezed in the radial direction. The evolution of the hole
wavefunction with increasing magnetic field is depicted in
Fig.~14. At $B=0T$ (Fig.~14(a)), the main part of the hole
wavefunction is situated at the radial boundary of the stacked
system and its probability to sit at the radial side is
$P_{side}=73\%.$ There is some extent of the wavefunction towards
the top and bottom of the stack, and also between the dots. When
the magnetic field increases (Fig.~14(b), $B=10T$), the
wavefunction is pushed further between the disks, because it
prefers to sit as closely as possible to the electron. However,
there is not enough space between the disks to confine the whole
wavefunction, and therefore it is energetically more favourable
for the hole to jump to a higher angular momentum state. Further
increasing the magnetic field leads to more angular momentum
transitions (Figs.~14(c-f)).

As can be seen from the wavefunctions, the ground state for this
system is always symmetric. The asymmetry, as found in the
two-disks system, is however still present, but it turns out not
to be the ground state. For the $l_{h}=0$ state, the wavefunctions
become asymmetric at about $B=20T$,resulting in a slight bow in
the full curve of Fig.~13. The fact that the asymmetry is less
pronounced in this system can be understood as follows: in the
three-disks system the main part of the electron is located in the
middle disk. This is in contrast to the two-disks system, where
the electron wavefunction has equal parts sitting in the two
disks. Even when an asymmetry is induced in the present
three-disks system, the electron will still be mainly located in
the middle disk, and therefore the hole will sit more tightly at
the radial boundary than was the case for the two coupled disks.

\subsection{Three vertically coupled dots:\ large interdot distance $d_{z}$}

When we increase the interdot distance $d_{z}$ up to $5.5nm$, the
hole will start to be confined between the disks. Fig.~15 shows
that there are no angular momentum transitions for this system and
it resembles very much the narrow disk-like system (see Fig.~2).
In this figure the solid curve denotes the $l_{h}=0$ state,
whereas the dashed curve denotes the $l_{h}=1$ curve. We turn
again to the wavefunctions to explain this behaviour. The electron
is sitting mainly in the middle dot, this time with smaller extent
into the upper and lower dots (Figs.~15(a) and (b) for
respectively $B=0T$ and $B=50T$). The contourplot for the hole
wavefunction (Fig.~16(a)) shows that already at $B=0T $ the hole
is completely situated between the disks. This is the preferred
place for the hole, as it tends to sit as close as possible to the
electron. Further increasing the magnetic field, squeezes the
wavefunction more in the radial direction (Fig.~16(b)). As there
is still enough space between the disks, there is no need for the
hole to jump to a higher angular momentum state, and $l_{h}=0$
remains the ground state over the whole $B$-region.

For this case, no asymmetry occurs up to $B=50T$. This means that
this system is quite stable against spontaneous symmetry breaking.
This stability is due to the fact that again the electron is
strongly located in the middle disk, and that the hole is now
sitting between the disks instead at the radial boundary.

\subsection{Revision of the results for the single disk systems}

We now turn back to the single disk systems, both the disk-like
and the pillar-like, where we did not talk about broken symmetry
solutions. However also in this case they can be found.

We find that the disk-like system, i.e. $R=10nm$ and $d=2nm$, is
perfectly stable up to $B=50T$ and we find no asymmetric
behaviour. This is in perfect agreement with the result for the
three coupled disks with $d_{z}=5.5nm$. Indeed, for the disk-like
system we have again the electron sitting in the disk with the
hole above and below the disks, which is quite similar to the
second three-disks system.

For the pillar-like system, i.e. $R=4nm$ and $d=12nm$, we do find
an asymmetry, but only for the $l_{h}=0$ state and at very high
magnetic fields, i.e. starting around $B \simeq 80T$. The
difference in energy between the states with symmetric and
antisymmetric wavefunctions is only $0.08meV$ for $B=90T$. The
ground state however remains symmetric over the total considered
$B$-region. Thus the wavefunctions, as plotted in Fig.~4, remain
unchanged.

\section{Conclusions}

We studied an exciton in a type II quantum disk of radius $R$ and
thickness $d.$ For the confinement potential we took a hard wall
of finite height, with the hole located in the barrier, which is
only confined by the interaction with the electron, confined
inside the dot. We calculated the exciton energy and wavefunction
using the Hartree-Fock mesh method, which allows us to start
without any knowledge of the single hole wavefunction. We limit
ourselves to the study of a {\it model} system in which strain
effects are neglected.

In the first part, we examined the case of an exciton in a single
disk. First we investigated the exciton properties in the absence
of a magnetic field. The calculation of the probability for the
hole wavefunction to sit at the radial boundary tells us that we
can distinguish two ''regimes'': disk-like systems with $d<<2R,$
where the hole will prefer to sit above and below the quantum
disk, and pillar-like systems with $d>>2R,$ where the hole will be
located at the radial boundary of the disk.

Applying a magnetic field along the $z$-direction results in a
different behaviour for these two systems. For the disk-like
system, the hole is squeezed in the radial direction by the
magnetic field, and the ground state of the system is the
$l_{h}=0$ state for all values of $B.$ For the pillar-like system,
the magnetic field pushes the hole closer to the disk boundary,
which forms a barrier for the hole. Therefore it will be
energetically more favourable for the hole to jump to a higher
angular momentum $l_{h}$ state, as then the wavefunction is able
to relax away from the radial boundary. With increasing magnetic
field, we find successive $l_{h}$ transitions.

The investigation of a system of two vertically coupled disks was
done in order to compare its magnetic field dependence with the
pillar-like single-disk system. As was expected, we find again
angular momentum transitions of the ground state. Moreover, a new
feature appeared in this study, namely a spontaneous symmetry
breaking, induced by both the Coulomb interaction and the magnetic
field. We found though that with increasing magnetic field, it is
still preferable for the hole to jump to a higher angular momentum
state. Investigation of a system of two vertically coupled
truncated cones learned us that also in this case angular momentum
transitions occur, although the system is highly asymmetric.

A system of three vertically coupled disks was investigated in the
third part of the paper. It was shown that the system with small
interdot distance $d_{z}$ behaves similarly to that of the single
pillar-like disk. Again we found angular momentum transitions with
increasing magnetic field. For a larger interdot distance however,
the hole wavefunction tends to sit between the disks. In this
case, an increasing magnetic field does not lead to angular
momentum transitions anymore, and the $l_{h}=0$ state remains the
ground state. No spontaneous symmetry broken states are found as
the ground state. This is similar to the single disk system where
symmetry broken states occur only as excited states.

\section{Acknowledgments}

K. L. J. is supported by the ``Instituut voor de aanmoediging van
Innovatie door Wetenschap en Technologie in Vlaanderen'' (IWT-Vl)
and B. P. is a post-doctoral researcher with the Flemish Science
Foundation (FWO-Vl.). Discussions with M. Hayne, M. T\u{a}di\'{c}
and A. Matulis are gratefully acknowledged. Part of this work was
supported by the FWO-Vl, The Belgian Interuniversity Attraction
Poles (IUAP), the Flemish Concerted Action (GOA) Programmes, the
University of Antwerp (VIS) and European Commission GROWTH
programme NANOMAT project, contract no. GSRD-CT-2001-00545.

\section{Appendix: Calculation of the Hartree integral}

The Hartree integral expresses the interaction effect of one
particle on another, and is given by
\begin{equation}
\int \frac{\rho (r^{\prime },z^{\prime })}{|{\bf r}-{\bf r}^{\prime }|}d{\bf %
r}^{\prime }=\int dz^{\prime }\int dr^{\prime }r^{\prime }\int d\varphi
^{\prime }\frac{\rho (r^{\prime },z^{\prime })}{\sqrt{r^{2}+r^{\prime
2}-2rr^{\prime }\cos (\varphi -\varphi ^{\prime })+\left( z-z^{\prime
}\right) ^{2}}},  \eqnum{A1}
\end{equation}
which contains a $z$-dependence which was absent in our previous
work (see Appendix of Ref.~\cite{karen2}). We can remove the
$\varphi $-dependence, since we deal with a cylindrical symmetric
system. The integral over the angle becomes the complete elliptic
integral of the first kind, which converts Eq.~(A1) into
\begin{equation}
4\int dz^{\prime }\int \frac{\rho (r^{\prime },z^{\prime })r^{\prime }}{%
\sqrt{\left( r+r^{\prime }\right) ^{2}+\left( z-z^{\prime }\right) ^{2}}}%
{\cal K}\left( \frac{4rr^{\prime }}{(r+r^{\prime })^{2}+\left( z-z^{\prime
}\right) ^{2}}\right) dr^{\prime }.  \eqnum{A2}
\end{equation}
The radial integral has to be solved numerically. We use a
polynomial approximation for the elliptic function
\cite{abramowitz}, namely
\begin{equation}
{\cal K}(x)=[a_{0}+a_{1}x^{\prime }+a_{2}x^{\prime 2}]-[b_{0}+b_{1}x^{\prime
}+b_{2}x^{\prime 2}]\ln (x^{\prime }),  \eqnum{A3}
\end{equation}
with $x^{\prime }=1-x$ and where the coefficients $a_{i}$ and
$b_{i}$ are given in Ref.~\cite{abramowitz}. The commonly used
trapezoidal rule will give bad results, as this elliptic function
implies the appearance of a logaritmic divergence in the
integrand. Therefore we introduced the so-called `logaritmically
weighted method' which takes care of this problem.

Generally, the following integral can be considered:
\begin{equation}
I(r,z)=\int_{0}^{1}dxF(x)\ln \left( \frac{(x-r)^{2}+\left( z-z^{\prime
}\right) ^{2}}{(x+r)^{2}+\left( z-z^{\prime }\right) ^{2}}\right) ,
\eqnum{A4}
\end{equation}
which, after transformation, becomes
\begin{equation}
I(r,z)=\sum_{i=0}^{N-1}\int_{0}^{h}dxF(x+hi)\ln \left( \frac{%
(x-(r-hi))^{2}+\left( z-z^{\prime }\right) ^{2}}{(x+(r+hi))^{2}+\left(
z-z^{\prime }\right) ^{2}}\right) ,  \eqnum{A5}
\end{equation}
with $h$ the discretization step and $N$ the number of steps. If
we replace $F(x+hi)$ by $F_{i}+[F_{i+1}-F_{i}]\left( x/h\right) ,$
we can write (A5) as
\begin{equation}
I(r,z)=\sum_{i=0}^{N-1}\{F_{i}A_{i}(r,z)+[F_{i+1}-F_{i}]C_{i}(r,z)\},
\eqnum{A6}
\end{equation}
and the remaining problem is the calculation of the coefficients
$A_{i}(r,z)$ and $C_{i}(r,z).$ The integrals which determine the
coefficients can be solved exactly, which leads to the following
results:
\begin{equation}
A_{i}(r,z)=a(h-(r-hi))-a(h+(r+hi)),  \eqnum{A7}
\end{equation}
with
\begin{eqnarray}
a(y) &=&P\left( y\right) +P\left( h-y\right) +2\left| z-z^{\prime }\right|
\left( \arctan \frac{y}{\left| z-z^{\prime }\right| }+\arctan \frac{h-y}{%
\left| z-z^{\prime }\right| }\right) ,  \eqnum{A8(a)} \\
P\left( y\right) &=&y\ln \left( y^{2}+\left( z-z^{\prime }\right)
^{2}\right) ,  \eqnum{A8(b)}
\end{eqnarray}
and
\begin{equation}
C_{i}(r,z)=h^{-1}[c(h-(r-hi))-c(h+(r+hi))]-2r,  \eqnum{A9}
\end{equation}
with
\begin{eqnarray}
c(y) &=&\frac{1}{2}\left( y\left( 2h-y\right) ^{2}+\left( z-z^{\prime
}\right) ^{2}\right) Q\left( y\right) +\frac{1}{2}\left( \left( h-y\right)
^{2}-\left( z-z^{\prime }\right) ^{2}\right) Q\left( h-y\right)  \nonumber \\
&&+2\left| z-z^{\prime }\right| \left( h-y\right) \left( \arctan \frac{y}{%
\left| z-z^{\prime }\right| }+\arctan \frac{h-y}{\left| z-z^{\prime }\right|
}\right) ,  \eqnum{A10(a)} \\
Q\left( y\right) &=&\ln \left( y^{2}+\left( z-z^{\prime }\right) ^{2}\right)
.  \eqnum{A10(b)}
\end{eqnarray}

\bigskip

\begin{figure}[tbp]
\caption{(a) Phase diagram of the probability for the hole
wavefunction to sit at the radial border of the disk, as a
function of both $R$ and $d.$ (b), (c) and (d) are contourplots of
the hole wavefunction, at the positions as indicated by the
corresponding numbers on the main figure. The dashed lines in the
contourplots indicate the boundary of the disk.}
\end{figure}

\begin{figure}[tbp]
\caption{The variation of the exciton energy with increasing
magnetic field, for a disk with thickness $d=2nm$ and radius
$R=10nm$ and two values of the hole angular momentum. The insets
show the electron (dashed curve) and hole (solid curve)
wavefunction for (a) $B=0T$ and (b) $B=50T.$}
\end{figure}

\begin{figure}[tbp]
\caption{Evolution of the exciton energy with increasing magnetic
field, for a disk with thickness $d=12nm$ and radius $R=4nm.$
Increasing the magnetic field, leads to successive transitions of
the hole angular momentum $l_{h}$ indicated by the arrows. The
inset depicts a countourplot of the electron wavefunction at
$B=0T.$ The dashed line indicates the disk boundary.}
\end{figure}

\begin{figure}[tbp]
\caption{Contourplots of the hole wavefunction for a disk with
$R=4nm $ and $d=12nm,$ at (a) $B=0T$, (b) $B=25T$, (c) $B=50T$ and
(d) $B=90T$. The dashed lines indicate the disk boundary.}
\end{figure}

\begin{figure}[tbp]
\caption{Phase diagram of the angular momentum transitions as a
function of the hole confinement potential $V_{h}$ and the
magnetic field $B,$ for a disk radius $R=6nm$ and thickness
$d=14nm.$}
\end{figure}

\begin{figure}[tbp]
\caption{Overlap integral as a function of the confinement
potential of the hole, for $B=0T$ (solid curve) and $B=25T$
(dashed curve).}
\end{figure}

\begin{figure}[tbp]
\caption{Exciton energy as function of the magnetic field, for two
coupled disks with $R=6nm$, $d=6nm$ and $d_{z}=3.6nm$. The insets
show contourplots of the wavefunctions at $B=0T$, for respectively
the electron (a) and the hole (b).}
\end{figure}

\begin{figure}[tbp]
\caption{Contourplots of the hole wavefunction for two coupled
disks, for $l_{h}=0$ at respectively $B=15T$(a), $B=30T$ (b) and
$B=50T$ (c) and for $l_{h}=3$ at $B=50T$ (d).}
\end{figure}

\begin{figure}[tbp]
\caption{Degree of asymmetry of the different $l_{h}$ states as a
function of the magnetic field. The inset shows the dipole moment
as a function of the magnetic field.}
\end{figure}

\begin{figure}[tbp]
\caption{ Contourplots of the electron (a) and hole (b)
wavefunction at $B=0T$ for two coupled disks with $R=12nm$,
$d=3nm$ and $d_{z}=3nm$.}
\end{figure}

\begin{figure}[tbp]
\caption{Evolution of the exciton energy with the magnetic field
for two coupled truncated cones. The insets show contourplots of
the electron (a) and hole (b) wavefunctions at $B=0T$. The dashed
lines indicate the disk boundary.}
\end{figure}

\begin{figure}[tbp]
\caption{Contourplots of the hole wavefunction at (a) $B=20T$, (b)
$B=30T$ and (c) $B=50T$ for a double dot system made out of
truncated cones.}
\end{figure}

\begin{figure}[tbp]
\caption{Exciton energy as a function of the magnetic field, for
three vertically coupled disks with interdot distance $d_{z}=3nm.$
Increasing the magnetic field leads to angular momentum
transitions, which are indicated by the arrows. The insets show
contourplots of the electron wavefunctions at respectively (a)
$B=0T$ and (b) $B=50T. $ The dashed lines indicate the disk
boundaries. }
\end{figure}

\begin{figure}[tbp]
\caption{Contourplots of the hole wavefunction for (a) $B=0T$, (b)
$B=10T,$ (c) $B=15T,$ (d) $B=25T,$ (e) $B=35T$ and (f) $B=50T.$
The dashed lines indicate the disk boundaries.}
\end{figure}

\begin{figure}[tbp]
\caption{Exciton energy as a function of the magnetic field, for
three vertically coupled disks with interdot distance
$d_{z}=5.5nm.$ The insets are contourplots of the electron
wavefunctions for (a) $B=0T$ and (b) $B=50T. $ The dashed lines
indicate the disk boundaries. }
\end{figure}

\begin{figure}[tbp]
\caption{Contourplots of the hole wavefunctions for (a) $B=0T$ and
(b) $B=50T.$ The dashed lines indicate the disk boundary. }
\end{figure}


\begin{references}
\bibitem{uitleg} We allowed for broken symmetry solutions by
manually pushing the electron single particle wavefunction 0.1\%
more to the positive $z$-direction with respect to the negative
$z$-direction in the first iteration step.
\end{references}
\end{document}